\documentclass[letter]{aa}  

\usepackage{graphicx}
\usepackage{txfonts}
\usepackage{lipsum}
\usepackage{ulem}
\usepackage{float}
\usepackage{stfloats}
\usepackage{afterpage}
\usepackage{threeparttable}
\usepackage{lscape}             
\usepackage{placeins}           
\usepackage[ 
 breaklinks=true,
  colorlinks=true,         
   urlcolor=pdfurlcolor,    
   filecolor=pdffilecolor,  
   linkcolor=pdflinkcolor,  
  citecolor=pdfcitecolor,  %
]{hyperref}

\begin{document}

\definecolor{pdfurlcolor}{rgb}{0,0,0.6}
\definecolor{pdffilecolor}{rgb}{0.7,0,0}
\definecolor{pdflinkcolor}{rgb}{0,0,0.6}
\definecolor{pdfcitecolor}{rgb}{0,0,0.6}

\newcommand{\xmm}{{\it XMM-Newton}}
\newcommand{\erosita}{{eROSITA}}
\newcommand{\eROSITA}{{eROSITA}}
\newcommand{\cncha}{{CN\,Cha}}
\newcommand{\einstein}{{\it EINSTEIN}}
\newcommand{\chandra}{{\it Chandra}}
\newcommand{\swift}{{\it Swift}}
\newcommand{\gaia}{{\it Gaia}}
\newcommand{\etal}{et al.}
\newcommand{\nh}{\mbox {$N_{\rm H}$}}
\newcommand{\sss}[1]{\textcolor{blue}{\textbf{Sara: #1}}}   

   \title{First X-ray detection of symbiotic nova CN Cha with eROSITA/SRG}

%
%
%

    \titlerunning{First X-ray detection of symbiotic nova \cncha\ by eROSITA}
   \authorrunning{S.~Saeedi et al}
   \author{Sara Saeedi\inst{1}, Jaroslav Merc\inst{2,3}, Manami Sasaki\inst{1}, Lorenzo Ducci\inst{4}, Michael Freyberg\inst{5}}
\vspace{-1.0 cm}
\institute{\\\inst{1}  Dr. Karl Remeis-Sternwarte, Erlangen Centre for Astroparticle Physics, Friedrich-Alexander-Universit\"at Erlangen-N\"urnberg, Sternwartstrasse 7, 96049, Bamberg, Germany\\
    \email{sara.saeedi@fau.de}\\
    \inst{2} Astronomical Institute of Charles University, V Hole\v{s}ovi\v{c}k{\'a}ch 2, Prague, 18000, Czech Republic\\
    \inst{3} Instituto de Astrof\'isica de Canarias, Calle Vía Láctea, s/n, E-38205 La Laguna, Tenerife, Spain\\
    \inst{4} Institut f\"ur Astronomie und Astrophysik, Universit\"at T\"ubingen, Sand 1, 72076 T\"ubingen, Germany\\
    \inst{5} Max-Planck-Institut f\"ur extraterrestrische Physik, Gießenbachstraße 1, 85748 Garching bei M\"unchen, Germany}
   \date{Received \today}

 
 \abstract
   {}
   {We report on the first X-ray detection of the symbiotic star \cncha\ with eROSITA\,(extended Roentgen Survey with an Imaging Telescope Array) onboard Spektrum-Roentgen Gamma\,(Spektr-RG). }
   {We analyse the eROSITA observations from eRASS2 to eRASS5 individually, as well as the combined eRASS:5 data of all five surveys in order to investigate the X-ray spectral and timing properties of \cncha\ throughout the eRASS. }
   {Although \cncha\ was not detected during the first eROSITA All-Sky Survey (eRASS1), it was detected for the first time in the second eRASS2. Our analysis shows that the X-ray emission of \cncha\ gradually  increased from eRASS2 to eRASS5. The X-ray spectrum of the source, with the main emission at $<$2.4\,keV, allows us to  classify it as a $\beta$-type symbiotic star. However, a more detailed analysis suggests that there is very soft component, which may originate either from thermonuclear burning on the white dwarf, or from the wind collisions. Additional X-ray observations are required to better understand the origin and emission mechanisms of the source.}
   {}


   \maketitle
   \nolinenumbers

\section{Introduction}
\cncha~(= Gaia~DR3~5199012973569013120; $\alpha_{\rm 2000}$\,= 10:59:57.63, $\delta_{\rm 2000}$\,=\,--79:57:01.07) is a relatively recently discovered Galactic symbiotic binary star \citep[see][for a recent review of symbiotic binaries]{2025Galax..13...49M} located at a distance of 3.06$\pm$0.23\,kpc \citep[][]{2021AJ....161..147B}. The object was originally catalogued as an apparently ordinary Mira variable \citep[][]{1963VeSon...6....1H}, exhibiting pulsations with a period of $\sim$250--260 d and an optical amplitude reaching $\sim$3 mag. Its symbiotic nature was revealed only after the onset of a symbiotic nova outburst in 2012/2013 \citep[][]{2020AJ....160..125L}, during which the system brightened by approximately 8 mag in the optical and has since been gradually returning toward quiescence. The eruption was also detected at millimeter wavelengths, with the mm brightening occurring with a delay relative to the optical rise \citep[][]{2026arXiv260501022T}.

\begin{figure*}[]
\includegraphics[width=\textwidth]{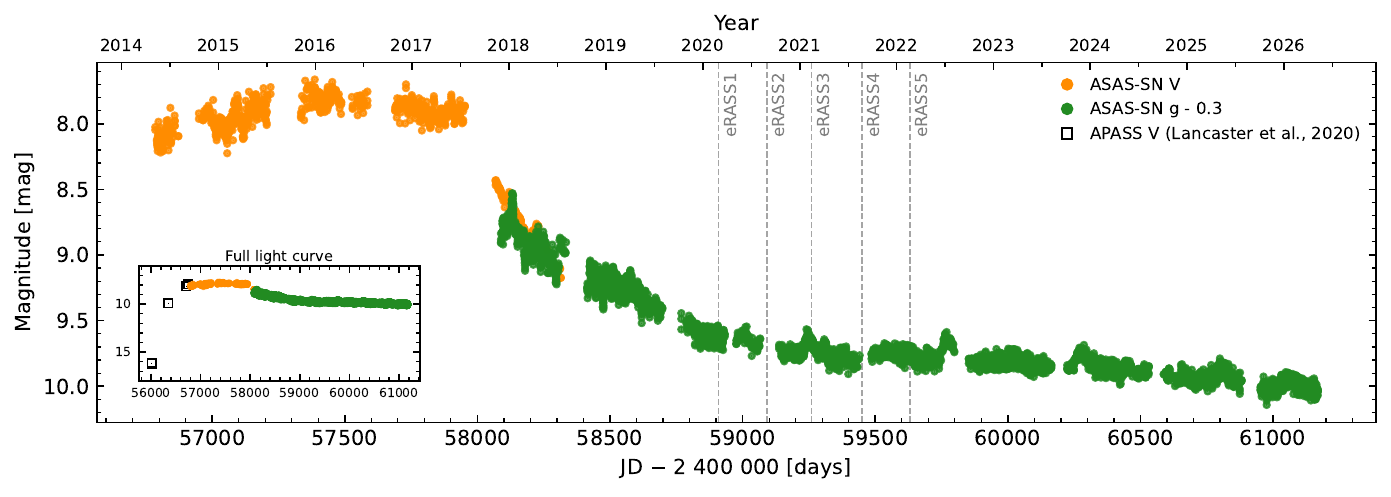}
\vspace{-0.5 cm}
\caption{Optical light curve of the symbiotic nova CN~Cha obtained from ASAS-SN $V$- and $g$-band photometry \citep[][]{2014ApJ...788...48S,2017PASP..129j4502K}. The $g$-band measurements are shifted by $-0.3$ mag for clarity. The vertical dashed lines mark the epochs of the eROSITA all-sky surveys eRASS1--eRASS5. The observations obtained before the outburst maximum are not shown in the main plot because of the large brightness range covered by the full light curve (see the inset).
  \label{fig:asas_sn_lc}}
\vspace{-0.1 cm}
\end{figure*}

In the optical, the eruption was characterised by a nearly flat maximum lasting for about three years, followed by a decline at a rate of $\sim$1.4 mag\,yr$^{-1}$ \citep[][]{2020AJ....160..125L}, as can be seen in the light curve in Fig. \ref{fig:asas_sn_lc}. Although \cncha\, belongs to the class of slow symbiotic novae, the duration of its plateau phase is among the shortest observed in such systems. Theoretical light-curve modelling based on hydrostatic envelope calculations indicates that the observed outburst properties can be reproduced by thermonuclear burning on a relatively low-mass white dwarf (WD, $\sim$0.6$\,M_\odot$) in a low-metallicity environment \citep[$Z = 0.004$;][]{2023ApJ...951..128K}.

The wide orbital separation of the binary\footnote{The orbital period of \cncha\, is unknown. However, orbital periods in symbiotic binaries typically range from a few years in S-type systems to several decades in (mostly D-type) systems with Mira donors, as \cncha.} likely played an important role in shaping the eruption morphology. In contrast to classical novae in cataclysmic variables, the Mira donor in \cncha\ remained well outside the expanding nova envelope, minimising dynamical perturbations and resulting in a smooth, single-peaked plateau rather than the complex multi-peaked behaviour often observed in more compact systems \citep[][]{2023ApJ...951..128K}.

Approximately 20 confirmed symbiotic stars listed in the New Online Database of Symbiotic Variables \citep[][]{2019RNAAS...3...28M,2019AN....340..598M,2026ApJS..285...45M} have undergone slow symbiotic nova outbursts. Roughly half of these systems have subsequently been detected in X-rays, although in many cases only decades after the nova eruption. Well-known examples include PU\,Vul, RR\,Tel, V1328\,Cyg, and AG\,Peg, the latter of which currently exhibits classical Z~And-type activity. Among distinguished classes of X-ray symbiotic stars \citep[see, e.g.,][]{1997A&A...319..201M,2013A&A...559A...6L}, most X-ray detected symbiotic novae are classified as $\beta$-type systems, which show a spectrum of a thermal plasma, consistent with emission from colliding winds and shocked circumstellar material with a pick of X-ray emission around 0.8\,keV \citep[][]{1997A&A...319..201M}. However, some objects are of $\alpha$-type with a very soft spectrum ($\lesssim$0.8\,keV), which is most likely emission from quasi-steady nuclear burning on the surface of a WD, while HM Sge exhibits a peculiar composite X-ray spectrum combining both $\alpha$- and $\beta$-type characteristics \citep[][]{2023ApJ...948...14T}.  Moreover, there are $\delta$-type symbiotic stars with hard X-ray emission\,($>$2.4~keV), where the emission is thought to originate from the boundary layer between the accretion disk and the white dwarf \citep[e.g.,][]{1997A&A...319..201M}. In many cases both soft and hard components have been observed in symbiotic stars ($\beta/\delta$ type), where one or two optically thin thermal plasma components are typically required to fit the spectra \citep[e.g.,][]{2013A&A...559A...6L}.

In this work, we report the first X-ray detection of \cncha\ with extended survey with an Imaging
telescope Array\,(eROSITA) onboard Spektrum-Roentgen Gamma \citep[SRG;][]{2021A&A...647A...1P}  and present an analysis of its currently available X-ray observations. Section~\ref{sect:data-redunction} provides the details of the X-ray data reduction and analysis, and in Section~\ref{sect:dis} we discuss the different possible scenarios for the X-ray emission mechanisms in \cncha.

\begin{figure*}[]
\includegraphics[trim={1.7cm 12.cm 1.2cm 13.cm},clip, width=1.0\textwidth]{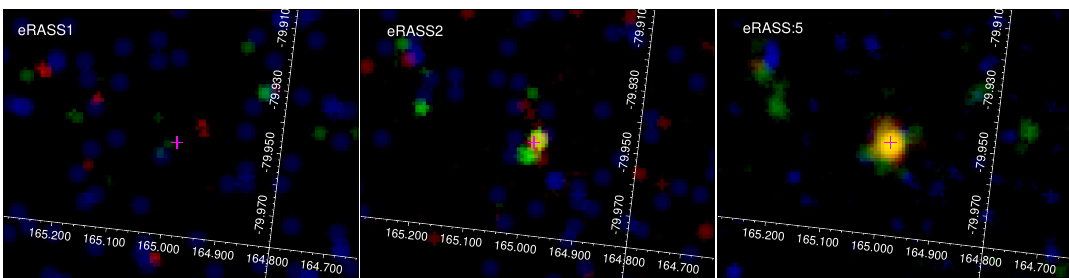}
\caption{The eROSITA detection of \cncha. The magenta cross indicates the position of the optical counterpart of the source. As shown in the \textbf{left panel} panel, no significant X-ray source is detected in the eRASS1 survey, while in eRASS2 \textbf{(middle panel)}, it is detected. \textbf{Right panel} shows the combined eRASS:5 image of the source. In these images, the red, green, and blue colours correspond to the 0.2--1.0 keV, 1.0--2.0 keV, and 2.0--8.0 keV energy bands, respectively.\label{fig:eROSITA-image} }
\vspace{-0.3 cm}
\end{figure*}  
\vspace{-0.3 cm}
\section{Data reduction and analysis}
\label{sect:data-redunction}
\subsection{X-ray Data}
We used individual eROSITA all sky survey\,(eRASS) data from eRASS1 to eRASS5, as well as the combined data from all five surveys (eRASS:5), to perform  spectral and timing analysis of \cncha. Data reduction and source detection were carried out using the eROSITA Science Analysis Software System\,(eSASS; Version:\,eSASSusers\_211214)\footnote{\url{https://erosita.mpe.mpg.de/edr/DataAnalysis/} }\citep{2018SPIE10699E..5GB,2021A&A...647A...1P}. For data reduction, source detection, and images, all seven TM cameras were used. However, for the timing and spectral analyses, data from cameras TM5 and TM7 were excluded due to light-leak contamination affecting these cameras \citep{2021A&A...647A...1P}.
\begin{figure}[]
\includegraphics[trim={0.cm 0.cm 0.0cm 0.cm},clip, width=0.44\textwidth]{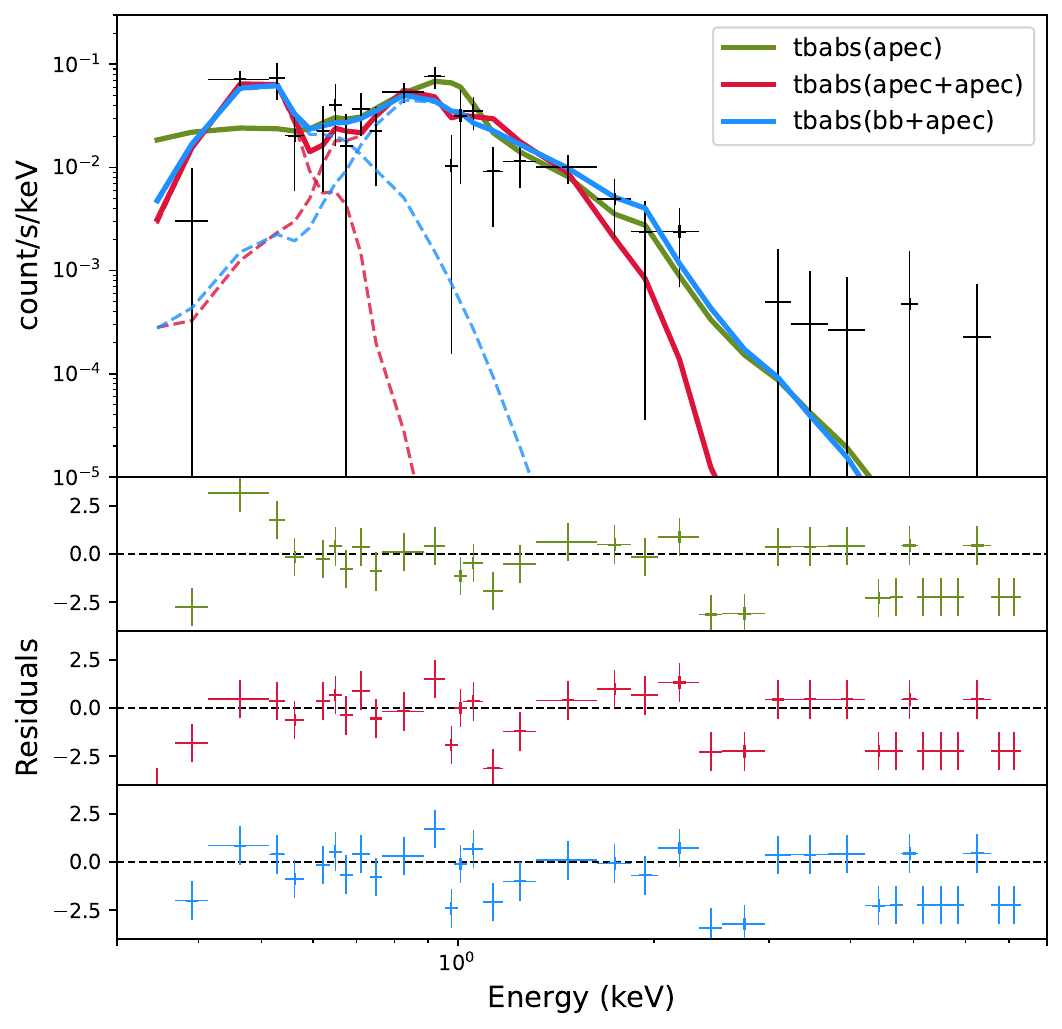}\\
\hspace*{3.3mm}\includegraphics[trim={0.cm 0.cm 0.0cm 0.cm},clip, width=0.366\textwidth]{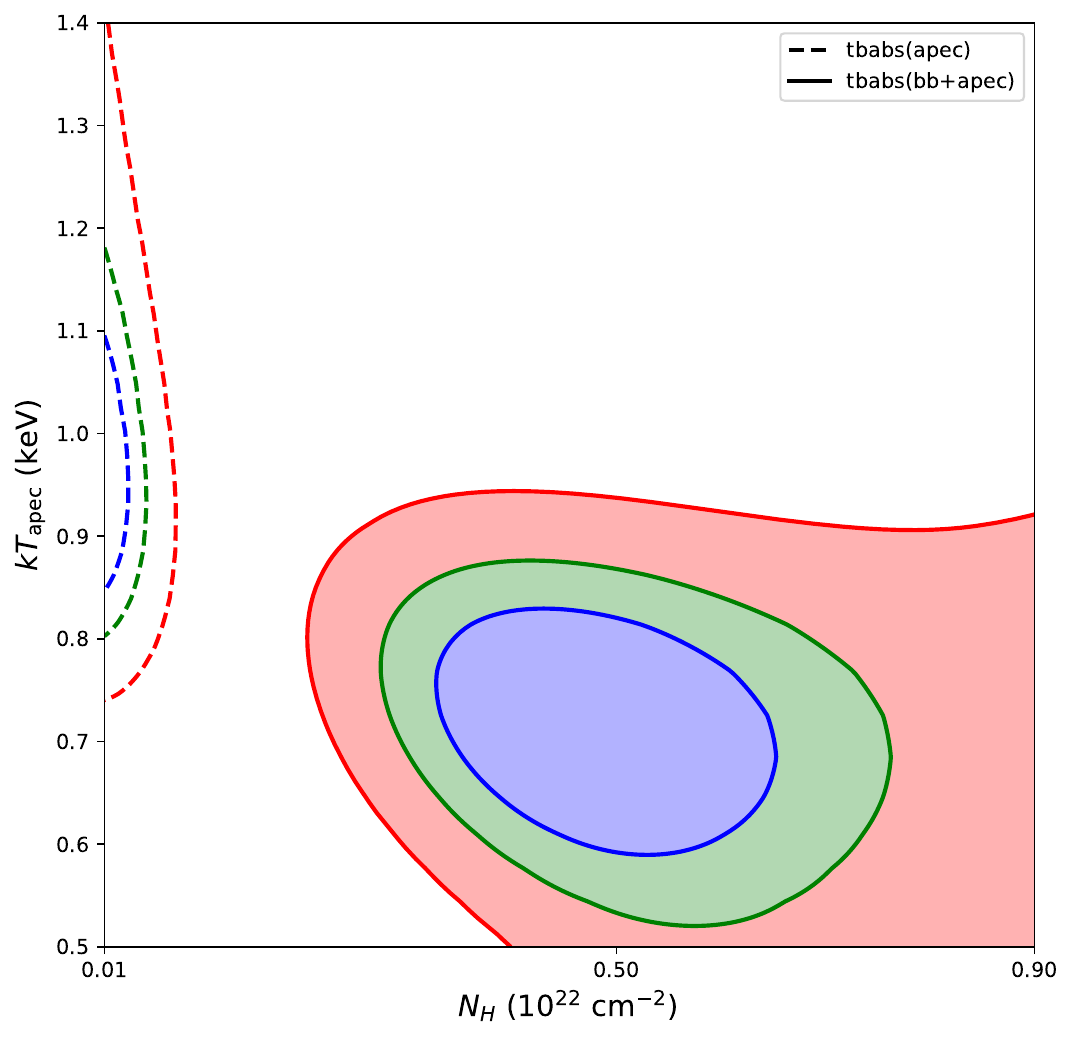}\hfill\\
\hspace{1cm}
\includegraphics[trim={0.cm 0.cm 0.cm 0.cm},clip, width=0.46\textwidth]{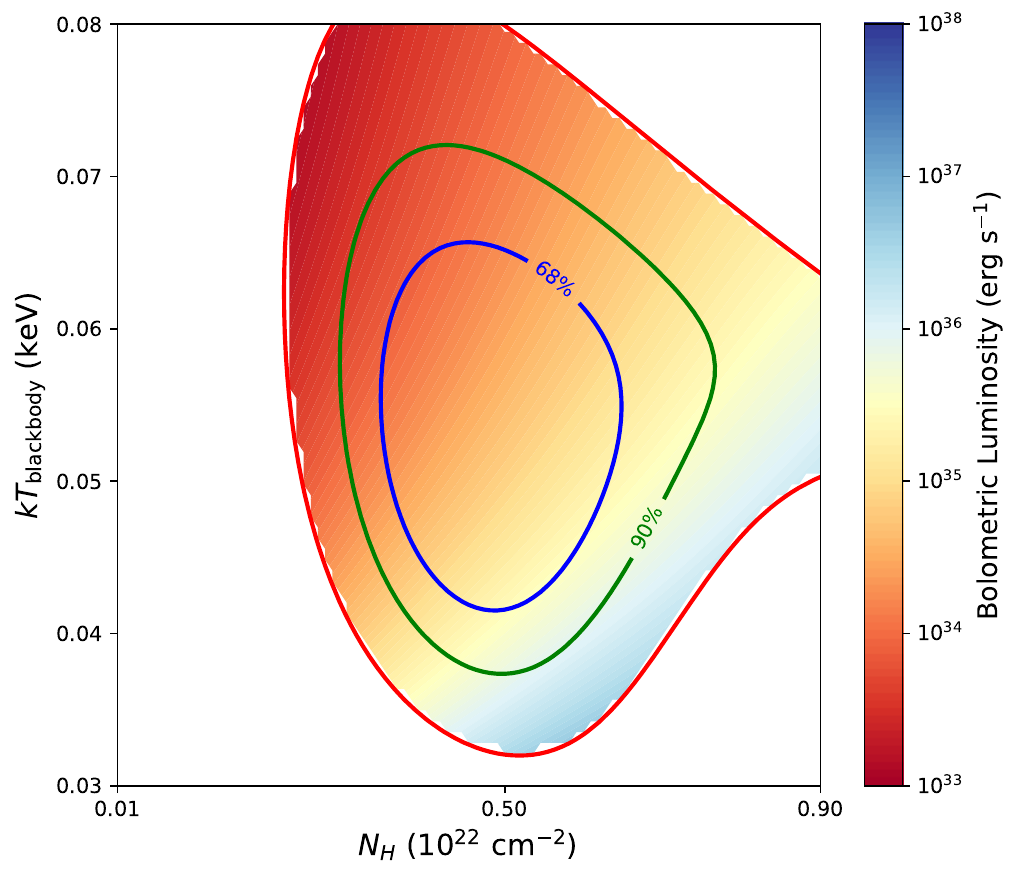}
\vspace{-0.3 cm}
\caption{\textbf{Upper:} Combined eROSITA eRASS:5 spectrum together with the residuals shown in units of standard deviation. The source spectrum is fitted with three different models. The solid lines represent the best-fit models, while the dashed lines indicate the individual model components. \textbf{Middle:} Contour plots of the \nh\ and temperature of the \texttt{apec} for the absorbed \texttt{apec} model (dashed lines) and the absorbed \texttt{blackbody+apec} model (solid lines) fitted to the spectrum. \textbf{Lower:} Contour plots of the \nh\ and temperature of the blackbody of absorbed \texttt{blackbody+apec} model. The colour map present the bolometric luminosity of blackbody model. The contours correspond to confidence levels of 68$\%$ (blue), 90$\%$ (green), and 99$\%$ (red). }
\label{fig:spec}
\end{figure}
\cncha\ was not significantly detected in eRASS1 and is therefore not reported in the eRASS1 catalogue \citep{2024A&A...682A..34M}. However, our analysis shows that \cncha, was detected in the eRASS2 \citep[]{2026A&A...712A.171R} and later. Figure~\ref{fig:eROSITA-image} shows the eROSITA images of \cncha\ from eRASS1, when the source was not yet detectable in X-rays; from eRASS2, when it was detected for the first time; and from the combined observation using all available eRASS data (eRASS:5). The details of each eRASS observation is available in Table.\,\ref{table:eROSITA}. 

\vspace{-0.1 cm}
In addition, since five archival \textit{Swift} \citep{2004SPIE.5165..169N} observations, obtained between 2007 and 2011, are available in the field of \cncha, we analysed data taken with \textit{Swift} X-ray Telescope \citep[XRT;][]{2005SSRv..120..165B} and the Ultraviolet/Optical Telescope \cite[UVOT;][]{2005SSRv..120...95R}. We analysed the XRT data taken in Photon Counting mode to derive X-ray upper limits for the detection of \cncha. We also extracted the magnitudes in all available UV and optical filters for each observation. Details of the \textit{Swift} observations, and the light curve are provided in Table~\ref{table:swift} and Fig.\,\ref{fig:light-curve-swift}. The results of the \textit{Swift} analysis are discussed in Sect.~\ref{sect:dis}. 
\vspace{-0.2 cm}
\subsection{Spectral and timing analyses} 
\vspace{-0.3 cm}
For the spectral analysis, we used the combined eRASS:5 data to obtain the best statistics. 
We tested different background regions to ensure that the extracted spectrum was not affected by the choice of background region. 
In our spectral analysis (using Xspec,\,version:\,12.15.1), we applied the W statistic\footnote{See: \url{https://heasarc.gsfc.nasa.gov/docs/software/xspec/manual/XSappendixStatistics.html}} implemented as a modified C-statistic in Xspec \citep{2009ApJ...693..822H}. Considering the low number of source counts due to the short eROSITA exposure times, we binned the spectrum using \texttt{ftgrouppha} to ensure a sufficient background contribution in each bin and to avoid underestimating the background during the spectral analysis, as recommended by \citet{2023hxga.book..150B} and \citet{2023arXiv230907577D}.

Following the classification of X-ray symbiotic stars by \citet{2013A&A...559A...6L}, $\beta$-type systems are symbiotic stars with emission peaking around 0.8 keV, typically below 2.4 keV, and \cncha\, shows, in general, similar emission. It is believed that in $\beta$-type symbiotic stars, the X-ray emission likely originates from ionised plasma, from the colliding winds of the  white dwarf in outburst and that of the red giant companion \citep{1997A&A...319..201M, 2013A&A...559A...6L}. Therefore, the X-ray emission of the system seems to suggest a $\beta$-type symbiotic star. From this reason, in the first step, we fitted an absorbed thermal plasma model to the source. We used the tbabs model \citep{2000ApJ...542..914W} for foreground absorption and collisionally ionized thermal plasma model \cite[apec,][]{2001ApJ...556L..91S} for the spectral analysis. Although the statistics is poor, a single \texttt{apec} model looks insufficient to provide an acceptable fit (see spectral fit in Fig.\,\ref{fig:spec} and Table \,\ref{table:spec}). Not only a soft excess is present <0.5\,keV, but also, the \texttt{tbabs(apec)} model suggests a very low absorption (see the middle plot in Fig.~\ref{fig:spec} and Table~\ref{table:spec}), even lower than the absorption in the direction of \cncha\, \citep[$6.53\times10^{20}\,\mathrm{cm}^{-2}$][]{2016A&A...594A.116H}. However, it is expected that the giant Mira companion, surrounded by a dusty circumstellar nebula produces noticeable intrinsic absorption. Therefore, this fit result indicates that an additional component is necessary to explain the X-ray emission of \cncha. We fitted the spectrum with two models: absorbed\,(\texttt{bb+apec}) and absorbed\,(\texttt{apec+apec}), using \texttt{tbabs} for foreground absorption again. In both cases, the fit was significantly improved. In Sect.\,\ref{sect:dis}, we discuss the plausible X-ray emission mechanisms for \cncha.
For each observation, the flux was calculated using the best-fit model for the source emission in that particular observation. For eRASS1, the upper-limit flux was estimated assuming an absorbed \texttt{apec} model with the parameters listed in Table~\ref{table:spec}.

The variability of the source across the five eRASS observations was investigated through both spectral and timing analyses. The sensitivity of eROSITA is limited to energies $<$\,2.3 keV, and the main emission from \cncha\ also lies below this energy. On the other hand, thermonuclear burning would cause emission only in the extremely soft band $\lesssim$0.8 keV. Therefore, we calculated the hardness ratio (HR) of the source using two energy bands: B1\,(0.2--0.8 keV) and B2\,(0.8--2.3 keV). Together with their corresponding uncertainties of energy bands, EB1 and EB2, the hardness ratio was calculated as $HR=(B2-B1)/(B2+B1)$, and its corresponding uncertainty was calculated using  $EHR=2\frac{\sqrt{(B2EB1)^2+(B1EB2)^2}} {(B2+B1)^2}$.  The results of flux and spectral variability over time is presented in Fig.\ref{fig:variability} and discussed in Sect.\ref{sect:dis}.
\section{Discussion}
\label{sect:dis}
The analysis of eROSITA data revealed that the source gradually brightened in X-rays from eRASS2 to eRASS5. As shown in the light curve in Fig.~\ref{fig:variability}, the flux in eRASS4 and eRASS5 is significantly higher than in eRASS2 and eRASS3. The upper limit from eRASS1 also indicates that the X-ray luminosity in that epoch was at least near one order of magnitude lower than in eRASS2, suggesting that the source changed its physical state between the two epochs.  As the HR diagram of the source in Fig. \ref{fig:variability} shows, the source was not detected in the softer band in eRASS2. While the luminosity was increasing in eRASS3 to eRASS5, the HR did not change significantly.

Looking into X-ray spectral studies of symbiotic stars, emission similar to \cncha\ has been reported, e.g., in HM~Sge \citep{2023ApJ...948...14T} and AG\,Peg \citep{2018MNRAS.481.5156Z}.
If we assume that the soft component of the emission comes from thermonuclear shell-burning present from the nova eruption on the surface of the white dwarf, the fitted absorbed\,(\texttt{bb+apec}) model suggests a temperature between 40--70\,eV\,[$(4.6-8.12)\times10^5$\,K] with an absorption varies between $(3.32-6.67)\times10^{21}$ (in 90\% confidence ranges). In this case, the increase in X-ray luminosity during the eRASSs, while the optical emission decreases, might be related to the geometry of the system. Over time, the optical depth may have decreased, allowing the soft X-ray emission to become visible.
Normalization ($K$) of the blackbody model provides the bolometric luminosity, $K =L_{39}/(D_{10}^{2})$, where $L_{39}$ is the source luminosity in units of $10^{39}$\,erg/s and $D_{10}$ is the distance of the source in units of 10\,kpc.
The bolometric luminosity of thermonuclear shell burning is expected to be in the range of $10^{36-38}$\,erg\,s$^{-1}$ \citep[e.g.][]{1997ARA&A..35...69K, 2016PhDT.......276S}. However, as shown in the  contour plot in the lower panel of Fig.~\ref{fig:spec}, this condition is satisfied only within a small region of parameter space of the column density and blackbody temperature outside the 90\% confidence region. Therefore, it is not possible with the currently available data to confirm the possibility that the soft emission component is produced by thermonuclear burning.

It could also be the case that the system may have a harder X-ray emission from the disk as it has been observed in case of e.g., T\,CrB, or  Y\,Gem \citep{2024MNRAS.528..987T}, V347\,Nor \citep{2013A&A...559A...6L}.
However, because of the low sensitivity of eROSITA for energies $>$2.3\,keV, we cannot detect any hard  emission.  If there is a disk, the very soft emission may originate from the boundary layer of disk in case of high-accretion, as the model of \citet{2024MNRAS.528..987T} suggests.  Also, if an accretion disk is present, the soft X-ray emission could potentially originate from the cooling of a wind from the disk.

On the other hand, there is also the possibility that the very soft component of the emission may originate not only from thermonuclear burning, but from shock-heated plasma in the winds, as observed in the case of RR\,Tel, which is also a slow symbiotic nova similar to \cncha\,\citep[][]{2013A&A...556A..85G}. In RR\,Tel, a detailed study of the temperatures of He-like triplet lines confirms that the super-soft component of the X-ray emission mainly originates from wind collision during the white dwarf outburst rather than from residual nuclear burning. 
Therefore, considering the various possible scenarios for the X-ray mission of \cncha, deeper X-ray observations are crucial to better constrain the model parameters, reduce the uncertainties, and provide a more accurate understanding of the origin and mechanisms of the X-ray emission from the source.

In addition, we have analysed the archival X-ray data from the \swift\ observatory covering the field of \cncha. The observation was carried out between 2007 to 2011 (Table\,\ref{table:swift}), which was before the optical outburst observed in 2012/2013 \citep{2020AJ....160..125L}.  A faint and variable UV emission is detected; however, in the X-ray band, only upper limits were obtained. These upper limits are comparable to the X-ray flux measured with eROSITA. Therefore, we cannot determine whether the source exhibited X-ray emission at a level similar to that observed during, for example, eRASS2, or whether it was in a fainter X-ray state during the \swift\ observations.

\begin{acknowledgements}
This work was supported by the Deutsche Forschungsgemeinschaft through the project SA 4388/2-1. The research of J.M. was supported by the Czech Science Foundation (GACR) project no. 24-10608O.
This work is based on data from
eROSITA, the soft X-ray instrument aboard SRG, a joint Russian-German science mission supported by the Russian Space Agency (Roskosmos), in the
interests of the Russian Academy of Sciences represented by its Space Research
Institute (IKI), and the Deutsches Zentrum für Luft- und Raumfahrt (DLR). The SRG spacecraft was built by Lavochkin Association (NPOL) and its subcontractors, and is operated by NPOL with support from the Max Planck Institute for Extraterrestrial Physics (MPE). The development and construction of the eROSITAX-ray instrument was led by MPE, with contributions from the Dr. Karl Remeis Observatory Bamberg \& ECAP (FAU ErlangenNuernberg), the University of Hamburg Observatory, the Leibniz Institute for Astrophysics Potsdam (AIP), and the Institute for Astronomy and Astrophysics of the University of Tübingen, with the support of DLR and the Max Planck Society. The Argelander Institute for Astronomy of the University of Bonn and the Ludwig Maximilians Universität Munich also participated in the science preparation for eROSITA. Also, We acknowledge the use of public data from the Swift data archive.
\end{acknowledgements}
\bibliographystyle{aa} 
\vspace{-0.5 cm}
\bibliography{bibtex}
\clearpage

\clearpage
\clearpage
\begin{appendix}

\begin{table*}[t]
\section{Tables and figures}
\centering
\begin{minipage}[t]{0.42\textwidth}
\centering
\caption{eROSITA observations in the field of \cncha\ (sky 
tile 172171).}
\label{table:eROSITA}
\begin{tabular}{lcc}
\hline\hline
OBS-ID & Date & Exp.Time\,(s)\\
\hline
eRASS1 & 2020-02-29;\,08:44 & 655\\
eRASS2 & 2020-08-29;\,18:14 & 665\\
eRASS3 & 2021-02-14;\,06:14 & 700\\
eRASS4 & 2021-08-23;\,12:14 & 653\\
eRASS5 & 2022-02-20;\,03:14 & 619\\
\hline
\end{tabular}
\end{minipage}
\hfill
\begin{minipage}[t]{0.54\textwidth}
\centering
\caption{Details of Swift observations in the field of \cncha.}
\label{table:swift}
\begin{tabular}{lccc}
\hline\hline
OBS-ID & Date & Exp.Time\,(ks) & UVOT Filters\\
\hline
00036766001 & 2007-12-22;\,02:19 & 2.80 & U\\
00036766002 & 2008-01-18;\,22:18 & 2.75 & U,\,UVW1\\
00036766003 & 2008-01-20;\,04:27 & 9.00 & UVM1,\,UVW2\\
00036766005 & 2010-08-30;\,06:17 & 0.66 & UVM2\\
00036766006 & 2011-08-26;\,00:54 & 2.50 & U,\,UVW1\\
\hline
\end{tabular}
\end{minipage}

\end{table*}

\begin{table*}
\caption{The parameters obtained from the spectral analysis using different spectral models.}   
\vspace{-0.2 cm}
\label{table:spec}     
\begin{tabular}{lccccccc}         
\hline\hline                      
Model & \nh &kT1 & kT2 &  C statistic& d.o.f &  Flux &Luminosity\\
      & 10$^{22}$cm$^{2}$ & keV & keV &  & &  erg\,s$^{-1}$\,cm$^{-2}$ & erg\,s$^{-1}$\\
\hline
tbabs(apec)& <0.03 &0.95$\pm$0.12&-- &66.05 & 33 &(1.27$\pm0.22)\times10^{-13}$ &(1.42$\pm0.50)\times10^{32}$\\[0.15cm]
tbabs(apec+apec)&  1.07$^{+0.23}_{-0.26}$&0.026$^{+0.004}_{-0.005}$&0.18$^{+0.09}_{-0.06}$ & 37.40& 31  &(1.01$\pm$0.17)$\times10^{-13}$ &(1.13$\pm0.40)\times10^{32}$\\[0.15cm]
tbabs(bb+apec)& 0.47$^{+0.19}_{-0.16}$&(bb)0.05$^{+0.02}_{-0.01}$&(apec)0.70$^{+0.15}_{-0.15}$& 35.63 &31 & (1.17$\pm$0.20)$\times10^{-13}$  &(1.31$\pm0.33)\times10^{32}$ \\[0.15cm]
\hline  
\end{tabular}
\begin{tablenotes}
      \small
      \item The abundance for all models was fixed to 0.2 $\times$ the solar abundance, consistent with the metallicity measurement reported by \citet{2023ApJ...951..128K}.
    \end{tablenotes}
\end{table*}
\begin{figure*}[t]
\includegraphics[trim={0.cm 0.cm 0.cm 0.cm},clip, width=0.4\textwidth]{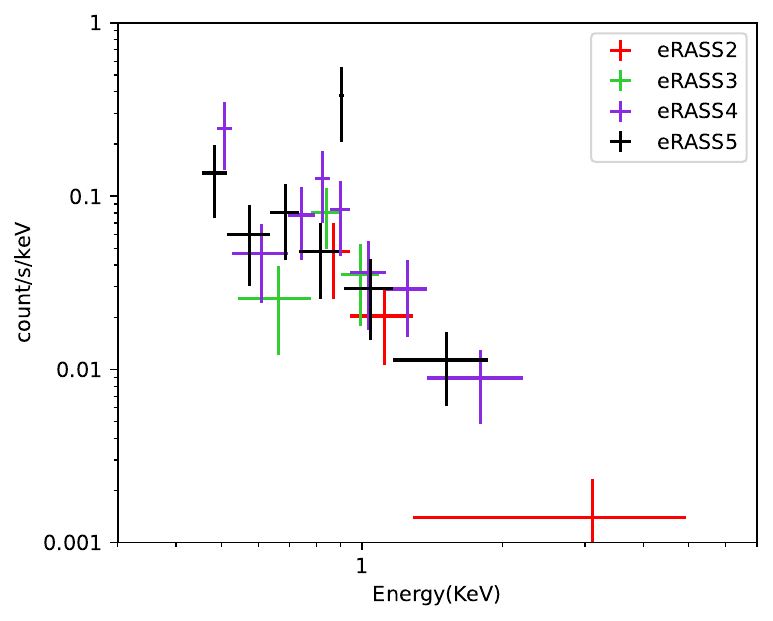}
\hspace{2. cm}
\includegraphics[trim={0.cm 0.cm 0.cm 0.cm},clip, width=0.4\textwidth]{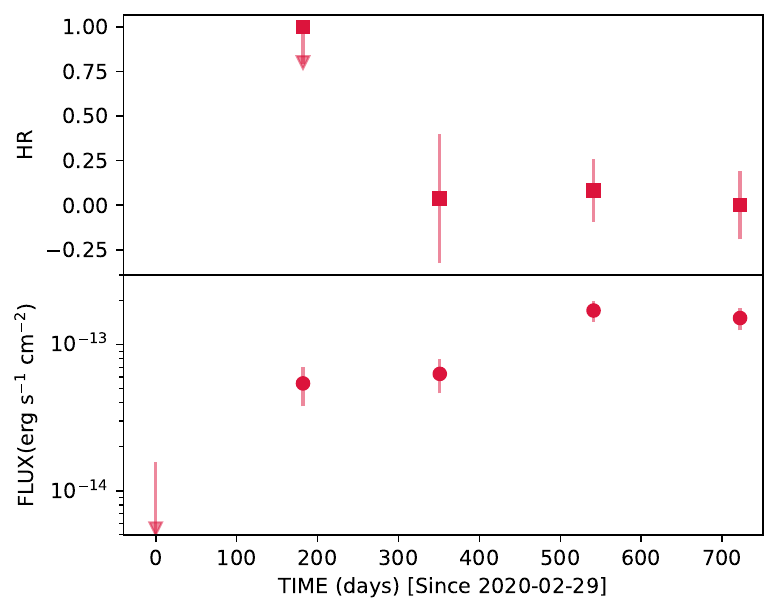}
\caption{\textbf{Left:} Spectra of the individual eRASS survey of \cncha. \textbf{Right:} The time evolution of the hardness ratio (HR) and flux variability of \cncha\ across the eRASS1 to eRASS5 surveys. The source was not detected in eRASS1; however, it becomes more luminous from eRASS2 to eRASS5. \label{fig:variability}}
\end{figure*}
\begin{figure}[!t]
\includegraphics[trim={0.cm 0.cm 0.cm 0.cm},clip, width=0.49\textwidth]{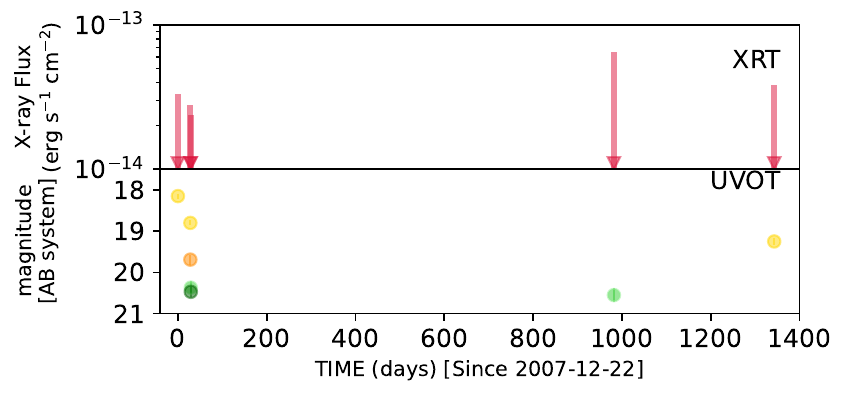}
\caption{ Light curves from \textit{Swift} spanning 2007 to 2011. The upper panel presents the X-ray upper limits at the position of \cncha. The lower panel shows the UVOT magnitudes of \cncha. Yellow, orange, light green, and dark green circles represent the $U$ (3465~\AA), $UVW1$ (2600~\AA), $UVM2$ (2246~\AA), and $UVW2$ (1928~\AA) filters, respectively.
\label{fig:light-curve-swift}}
\end{figure}
\FloatBarrier
\end{appendix}
\end{document}